\documentclass[aps,prd,groupedaddress,showpacs,showkeys]{revtex4}
\usepackage{latexsym,epsfig,amsmath,amssymb}
\usepackage{amssymb}
\usepackage{amsmath}
\usepackage{amsfonts}
\usepackage{epsfig}
\usepackage{verbatim}
\usepackage{feynmf}
\usepackage{graphicx}
\usepackage{float}
\newcommand{\seq}{\begin{subequations}}
\newcommand{\sen}{\end{subequations}}
\newcommand{\eq}{\begin{eqnarray}}
\newcommand{\en}{\end{eqnarray}}

\def\bwt{\begin{widetext}}
\def\ewt{\end{widetext}}
\def\be{\begin{equation}}
\def\ee{\end{equation}}
\def\bea{\begin{eqnarray}}
\def\eea{\end{eqnarray}}
\def\bean{\begin{eqnarray*}}
\def\eean{\end{eqnarray*}}
\def\bary{\begin{array}}
\def\eary{\end{array}}

\def\bit{\begin{itemize}}
\def\eit{\end{itemize}}

\def\lan{\langle}
\def\ran{\rangle}

\begin{document}

\title{Hyperon forward spin polarizabilty $\gamma_0$}

\author{
K.B. Vijaya Kumar$^1$,
Amand Faessler$^2$,
Thomas Gutsche$^2$,
Barry R. Holstein$^3$,
Valery E. Lyubovitskij$^2$
\footnote{On leave of absence from the
Department of Physics, Tomsk State University,
634050 Tomsk, Russia}
\vspace*{1.2\baselineskip}}

\affiliation{
$^1$
Department of Physics, Mangalore University,
Mangalagangothri, Karnataka 574199, India
\vspace*{1.2\baselineskip}
\\
$^2$
Institut f\"ur Theoretische Physik,
Universit\"at T\"ubingen,
Kepler Center for Astro and Particle Physics,
Auf der Morgenstelle 14, D--72076 T\"ubingen, Germany
\vspace*{1.2\baselineskip}
\\
$^3$
Department of Physics--LGRT, University of
Massachusetts, Amherst, MA 01003 USA
\vspace*{1.2\baselineskip}}

\date{\today}

\begin{abstract}
We present the results of a systematic leading order calculation of
hyperon Compton scattering and extract the forward spin
polarizability---$\gamma_{0}$---of hyperons within the framework of
SU(3) heavy baryon chiral perturbation theory (HBChPT). The results
obtained for $\gamma_{0}$ in the case of nucleons agree with that of
the known results of SU(2) HBChPT when kaon loops are not
considered.

\end{abstract}



\maketitle

\section{Introduction}

Compton scattering is a source of valuable information about baryons
since it offers access to some of the more subtle aspects of baryon
structure such as polarizabilities~\cite{AMB60}-\cite{BWF02}, which
parameterize the response of the target to an external quasi-static
electromagnetic field.  For the case of unpolarized nucleons the
spin-independent (SI) Compton amplitude is given by
\eq\label{Amp_SI}
\epsilon^{\mu}_{1} \, {\cal M}_{\mu\nu}^{\rm SI}
\, \epsilon^{\nu}_{2}
 = \vec{\epsilon} \cdot \vec{\epsilon}^{\, *}
\biggl( - \frac{Q^2_N}{m_N} + 4 \pi \alpha_{N} \, \omega \omega^{'}
\biggr) + 4 \pi \beta_{N} \ (\vec{\epsilon} \times \vec{q}) \cdot
(\vec{\epsilon}^{\, *} \times \vec{q}^{\,'}) \ + \ {\cal O}(\omega^{4})
\en
where $N = p, n$; $Q_N$, $m_N$ represent the nucleon charge and
mass, while $\epsilon_\mu = (0,\vec{\epsilon})$,
$\epsilon^{\, *}_\mu = (0,\vec{\epsilon}^{\, *})$ and 
$q_{\mu}=(\omega,\vec{q})$,
$q_{\mu}^{'}=(\omega^{'},\vec{q}^{\,'})$ specify the polarization
vectors and four-momenta of the initial and final photons,
respectively.  At this order the Compton amplitude is defined in
terms of two polarizabilities---electric ($\alpha_{N}$) and
magnetic ($\beta_{N}$), which measure the response of the nucleon to
applied quasi-static electric and magnetic fields.  By measurement of
the differential cross section one can extract $\alpha_{N}$ and
$\beta_{N}$ provided the energy is large enough such that the second
and third term in Eq.~(\ref{Amp_SI}) contribute significantly with
respect to the leading Thomson contribution, but is not so large that
higher order effects become significant.  This extraction has been
achieved in the energy range $50$ MeV $< \omega < 100$ MeV---for
a recent review see e.g. Refs.~\cite{DD03,MS05,PDV11}. According to
the Particle Data Group~\cite{PDG} current experimental numbers for
$\alpha_{N}$ and $\beta_{N}$ are: \eq & &\alpha_{p} = (12.0 \pm 0.6)
\times 10^{-4} \ {\rm fm}^3\,, \quad
    \beta_{p} = (1.9 \pm 0.5) \times 10^{-4} \ {\rm fm}^3\,, \nonumber\\
& &\alpha_{n} = (11.6 \pm 1.5) \times 10^{-4} \ {\rm fm}^3\,, \quad
    \beta_{n} = (3.7 \pm 2.0) \times 10^{-4} \ {\rm fm}^3\,.
\en The nucleon polarizabilities have been studied via a number of
theoretical approaches based on dispersion
relations~\cite{WP74,IG78,AIL81,AIL97,DD99,BP07,BRH94},
phenomenological Lagrangians~\cite{VP95,OS96,TF99,SK01,KR01},
constituent quark models~\cite{SC92,HL95,TUB05}, chiral-soliton type
of models~\cite{MC87,NNS89,SS92,WB93,NNS96} and lattice QCD using
the external electromagnetic field method in
quenched~\cite{FXL06,WD06} and unquenched approximation~\cite{ME07}.
Additional insights into the polarizabilities have come from chiral
perturbation theory (ChPT), an effective theory of the low-energy
strong interaction~\cite{SW79,JG84}, specifically from heavy baryon
chiral perturbation theory (HBChPT) which is an extension of ChPT
that includes the nucleon~\cite{EJ91,EJ92}.  The first such
calculations of nucleon polarizabilities within ChPT were carried out
in~\cite{BER91,BER92ONE}.  However, HBChPT has an important
deficiency in that the chiral perturbative series fails to converge
in part of the low energy region.  The problem is generated by a set
of higher order graphs involving insertions in nucleon lines.  It
has been shown that infrared singularities of the various one loop
graphs occurring in the chiral perturbation series can be extracted
in a relativistically invariant fashion.  This procedure is known as
infrared dimensional regularization (IDR)~\cite{TB99}. The IDR
respects the constraints of chiral symmetry as expressed through the
chiral Ward identities. The manifestly Lorentz-invariant form of
baryon chiral perturbation theory (BChPT) with the IDR prescription
has been successfully applied to calculate $\alpha_{N}$ and
$\beta_{N}$ and the results for these polarizabilities differ
substantially from the corresponding HBChPT
numbers~\cite{VL09,VL10}.  In addition, HBChPT has been employed to
analyze virtual Compton scattering processes since, as an
effective field theory, it satisfies the structures of gauge
invariance, Lorentz invariance and crossing symmetry~\cite{DD98}.
New predictions for generalized polarizabilities have been made
using HBChPT at $O(p^4)$ (NLO)~\cite{TRH097,TRHT98,TRH00} and, using
ChPT, Compton scattering from the deuteron has been computed to
order $O(p^4)$~\cite{SRB005}.  However, the situation with regard to
scattering from polarized targets is less satisfactory, in part
because few direct measurements of polarized Compton scattering have
been attempted.

The spin-dependent (SD) piece of the forward scattering amplitude
for real photons of energy $\omega$ and momentum $q$
is~\cite{BER95,ACH62,TRH98,BRH98,SRB05}, \eq \epsilon^\mu_1 \, {\cal
M}_{\mu\nu}^{\rm SD} \, \epsilon^\nu_2 = i e^2 \, \omega \,
W^{(1)}(\omega) \ \vec{\sigma} \cdot (\vec{\epsilon}
\times\vec{\epsilon}^{\, *}) + \ldots \en From the  theoretical
perspective there is particular interest in the low energy limit of
the amplitude: \eq e^2 W^{(1)}(\omega) = 4 \, \pi \, (f_2(0) +
\omega^2 \, \gamma_0^N) + \ldots \en where $\gamma_0$ is the forward
spin polarizability, which is related to the photo-absorption cross
sections for parallel $(\sigma_+)$ and antiparallel $(\sigma_-)$
photon and target helicities via
\eq \gamma_0^N = \frac{1}{4\pi^2}
\, \int\limits_W^\infty \frac{ds}{s^3}
 \, \Big[ \sigma_-(s) - \sigma_+(s) \Big] \,,
\en
where $W = M_{\pi} + M_{\pi}^2/(2 m_N)$ is the threshold energy
for an associated neutral pion in the intermediate state.
The Low-Gell-Mann-Goldberger low-energy theorem states that,
\eq
f_{2}(0)=-\, \frac{\alpha \, \kappa_N^2}{2 \, m^2_N} \,,
\en
where $\alpha=e^2/(4\pi) = 1/137.036$ is the fine-structure constant,
$\kappa_N$ is the nucleon anomalous magnetic moment~\cite{FLOW54}.

The forward spin polarizability $\gamma_{0}^N$ has been calculated
to $O(p^{3})$ (LO)~\cite{BER92} in the framework of HBChPT yielding,
at lowest order in the chiral expansion, \eq
\gamma_{0}^N=\frac{\alpha \, g^2_A}{24 \, \pi^2 \,  F^2 \, M_\pi^2}
= 4.54 \times 10^{-4} \ {\rm fm}^4 \en both for protons and
neutrons, where the entire contribution comes from $\pi N$ loops.
(Hereafter we shall use units of $10^{-4}\, {\rm fm}^4$ for the spin
polarizability).  This LO calculation of spin polarizability is a
{\it prediction}, since any low energy constants associated with the
polarizability enter only at next to leading order (NLO).  
At LO the polarizability
is given entirely by the loop contribution in terms of well known
parameters such as nucleon and pion masses and the pion-nucleon
coupling constant $(g_{\pi NN})$.  The effect of including the
$\Delta(1236)$ enters in counterterms at fifth order in standard
HBChPT, and has been estimated to be so large as to change the sign.
The forward nucleon spin polarizability $\gamma_{0}$ has been
computed in an extension of HBChPT with an explicit $\Delta$
in~\cite{TRH98}.

This calculation has also been carried out to NLO in the framework
of HBChPT~\cite{KBV2000,XJ2000,GCG2000,MCB01}. The contribution to
$\gamma_{0}^N$ up to and including NLO contributions is found to be
$\gamma_{0}^{p/n}=4.5-(6.9 \pm 1.5)$---the NLO contributions are
large.  The corresponding relativistic chiral one loop calculation
of the forward spin polarizability was carried out by Bernard et
al.~\cite{BER92} and the computed value of $\gamma_{0}^N$ was found
to be smaller than the LO result of HBChPT. The generalized
$\gamma_{0}^N$ has been calculated in the Lorentz invariant
formulation of BChPT to NLO which demonstrates a large NNLO
contribution~\cite{BER03,CWK06}. In~\cite{BER03} the quoted values
are $\gamma_{0}^p=4.64 $ and $\gamma_{0}^n =1.82$; hence the chiral
expansion does not seem to converge, which is attributed to the Born terms.
Also, as has been shown in Ref.~\cite{BER03}, inclusion of the
Born terms up to fourth order is not sufficient to obtain convergence and
thus a complete fifth order calculation seems mandatory.
However, when only the first two terms of the chiral expansion are
considered $(O(\mu^{-1}))$ the results reproduce the NLO HBCHPT results.
Electroproduction data have been used to extract $\gamma_{0}^N$
using the sum rule given above.
In particular, in Ref.~\cite{AMS94} the values $\gamma^{p}_{0}=-1.3$
and $\gamma^{n}_{0}=-0.4$ were found, while the analysis of
Ref.~\cite{DD98one} gives a smaller absolute values with
$\gamma^{p}_{0}=-0.6$ and  $\gamma^{n}_{0}=+0.0$ based
on the HDT parametrization.  The latest numerical results of
Schumacher~\cite{SCH2011,SCH2009} based on the photo production cross section
are $\gamma^{p}_0=-0.58 \pm 0.20$ and $\gamma^{n}_0=-0.38 \pm 0.22$.
The most recent results are
$\gamma^{p}_0=-0.90 \pm 0.08 \pm 0.11$~\cite{PAS2010}.
Other results based on different photomeson analyses are
$\gamma^{p}_0=-0.67$ (HDT), -0.65 (MAID), -0.86 (SAID) and -0.76 (DMT).
Hence it is safe to say that, although considerable progress has been made in
understanding $\gamma_{0}$ for the nucleon, the results obtained from
BCHPT/HBCHPT are far from the numerical results obtained from
the electroproduction data.
While a rather large amount of work has been devoted, both
theoretically and experimentally, to the study of the nucleon
polarizabilities, very little is known about {\it hyperon}
polarizabilities.  However, with the advent of hyperon beams at FNAL
and CERN, the experimental situation is likely to change, and this
possibility has triggered a number of theoretical investigations.
Already, predictions for electric and magnetic polarizabilities have
been made for low-lying octet baryons in the framework of  
LO HBChPT~\cite{BER92TWO}, and in the context of several other 
models, yielding a broad spectrum of predictions~\cite{PET81,LIP92,GOB96,%
NIS98,TAN2000,AAL11}.  At present, no experimental data is
available for the forward spin polarizabilty of the hyperons and no
theoretical calculations have been published.  Motivated by this
situation, in the present work we extend the analysis of SU(2)
HBChPT to the SU(3) version in order to compute $\gamma_{0}$ for
hyperons.  This could serve as a test of low-energy structure of QCD
in the three flavor sector.  However, there is also a need to compute
the spin polarizabilities in the framework of BChPT with the IDR
prescription. 

The paper is organized as follows. Section II contains an overview
of the SU(3) version of HBChPT relevant for the calculation of the hyperon
forward spin polarizabilities $\gamma_0$. The relevant Feynman rules
for the case of the $\Sigma^{+}$ polarizability are listed in
Appendix A (see Fig.1), and the required loop integrals are listed in Appendix
B. The explicit expressions for $\Sigma^{+}{\pi}^{+}(K^{+})$ loops
in terms of loop integrals are listed in Appendix C. In Section III
we give the explicit results for the hyperon spin polarizabilities
$\gamma_{0}$ and discuss the corresponding numerical results.
Brief conclusions are given in Section IV.

\section{Effective Lagrangian}

The lowest-order SU(3) HBChPT Lagrangian involving the octet
of pseudoscalar mesons~$\phi$
\eq
\phi={\sqrt{ 2}}\left(\begin{array}{ccc}
\frac{1}{\sqrt 2}\pi^0+\frac{1}{\sqrt 6}\eta &\pi^+&K^+\\
\pi^-&-\frac{1}{\sqrt 2}\pi^0+\frac{1}{\sqrt 6}\eta&K^0\\
K^-&\bar{K}^0&-\frac{2}{\sqrt 6}\eta\\
\end{array}\right)
\en
and the baryon octet $B$
\eq
B=\left(\begin{array}{ccc}
\frac{1}{\sqrt 2}\Sigma^0+\frac{1}{\sqrt 6}\Lambda &\Sigma^+&p\\
\Sigma^-&-\frac{1}{\sqrt 2}\Sigma^0+\frac{1}{\sqrt 6}\Lambda&n\\
\Xi^-&\Xi^0&-\frac{2}{\sqrt 6}\Lambda\\
\end{array}\right)
\en
consists of two basic pieces: the lowest-order chiral effective
meson Lagrangian  ${\cal L}^{(2)}_{\phi\phi}$~\cite{SW79,JG84}
\eq
{\cal L}_{\phi \phi}^{(2)}&=& \frac{F^2}{4} \, \lan \nabla_\mu U \,
\nabla^\mu U^\dagger \, + \, \chi_+ \ran
\en
and the lowest-order
meson-baryon Lagrangian ${\cal L}^{(1)\ {\rm
HHChPT}}$~\cite{EJ91,EJ92,BER95}: \eq {\cal L}^{(1)\ {\rm
HBChPT}}_{\phi \, B} = \lan\bar B (i \,v \cdot D) B\ran +
\frac{D}{F_0} \, \lan\bar B \,S^{\mu}\{u_{\mu}, B\}\ran +
\frac{F}{F_0} \, \lan\bar B \,S^{\mu} [u_{\mu}, B]\ran \,. \en where
the superscript $(i)$ attached to the above Lagrangians denotes their
low-energy dimension and the symbols $\lan \, \, \ran$, $[ \, \, ]$,
$\{ \, \, \}$ denote the trace over flavor matrices, commutator and
anticommutator, respectively.  We use the following notations: $U =
u^2 = \exp(i\phi/F_0)$, where $F_0$ is the octet decay constant (in
our calculations we use $F_0 = F_\pi = 92 $ MeV), $u_\mu = i
\{u^\dagger, \nabla_\mu u\}$; $\nabla_\mu$ and $D_\mu$ are the
covariant derivatives acting on the chiral and baryon fields,
respectively, including external vector $(v_\mu)$ and axial
$(a_\mu)$ fields:
\eq
\nabla_\mu U &=& \partial_\mu U - i (v_\mu +
a_\mu) U + i U (v_\mu-a_\mu)\,,
\nonumber\\
D_\mu B &=& \partial_\mu B + [\Gamma_\mu, B]
\en
with $\Gamma_\mu$
being the chiral connection given by
\eq
\Gamma_\mu=\frac{1}{2}[u^{\dagger},\partial_{\mu}u]
-\frac{i}{2}u^{\dagger}(v_{\mu}+a_{\mu})u
-\frac{i}{2}u(v_{\mu}-a_{\mu})u^{\dagger} \,.
\en
The covariant spin operator is
$S_{\mu}=\frac{i}{2}\,\gamma_{5}\,\sigma_{\mu \nu}v^{\nu}$,
obeying the following relations in $d$ dimensions~\cite{BER95}:
\eq
S \cdot v=0\,, \ \ S^{2}=\frac{d -
1}{4}\,, \ \ \{ S_{\mu},S_{\nu} \}=\frac{1}{2}(v_{\mu}
\,v_{\nu}-g_{\mu \nu})\,, \ \ [S_{\mu},S_{\nu}] = i \,\epsilon_{\mu
\nu \alpha \beta}\; v^{\alpha} S^{\beta} \,.
\en
Finally,
$\chi_\pm = u^\dagger \chi u^\dagger \pm u \chi^\dagger u$ with $\chi = 2 B
{\cal M} + \ldots$, where $B = |\lan 0|\bar q q |0 \ran|/F^2$ is the
quark vacuum condensate parameter and ${\cal M} = {\rm diag}\{ \hat
m, \hat m, \hat m_s \}$ is the mass matrix of current quarks (We
work in the isospin symmetry limit with $\hat m_{u}= \hat
m_{d}=\hat{m}=7$~MeV. The mass of the strange quark $\hat m_s$ is
related to the nonstrange one via $\hat m_s \simeq 25 \, \hat m$).
The parameters $D$ and $F$ are fixed from hyperon semileptonic
decays to be $D=0.80$ and $F=0.46$ with $D+F=g_{A}=1.26$ being the
nucleon axial charge. In the above equations, $m$ denotes the
average baryon mass in the chiral limit.

\section{Forward spin polarizability $\gamma_0$}

In order to calculate the forward spin polarizabilities, we work in
the Breit frame wherein the sum of the incoming and outgoing baryon
three-momenta vanishes.  We utilize the Weyl (temporal) gauge
$A_0=0$, which, in the language of HBChPT, means $v \cdot \epsilon
=0$, where $v_\mu = (1,0,0,0)$ is the baryon four-velocity. At
$O(p^3)$ only the loop diagrams contribute to $\gamma_0$---to one
loop, the hyperon polarizabilities are pure loop effects.  At LO
these loop diagrams have insertions only from ${\cal L}^{(1) \ {\rm
HBChPT}}_{{\phi}B}$.  Fig.~2 shows all the possible loop-diagrams,
which contribute to $\gamma_0$ for $\Sigma^+$.  Similarly for the
other octet baryons the diagrams in Fig.~2 are the only ones which
contribute to $\gamma_{0}$ (except that the incoming and outgoing
particles are different).  There do exist contact term graphs
stemming from two insertions from ${\cal L}^{(2) \ {\rm
HBChPT}}_{{\phi}\,B}$ and a single  insertion from ${\cal L}^{(3) \
{\rm HBChPT}}_{{\phi}\,B}$, but these do not contribute to
$\gamma_{0}$ and consequently we have not shown these diagrams in our
manuscript.  Appendix A (see Fig.1) lists the relevant Feynman rules for the
computation of the loop diagrams, while Appendix B contains the
relevant loop integrals required for their evaluation.  Appendix C
gives the analytic results for $\Sigma^{+}\pi^{+}(K^{+})$ loops
contributing to the forward Compton scattering amplitude $\gamma
\Sigma^+ \to \gamma \Sigma^+$.  Note that both pion and kaon loops
yield finite contributions to $\gamma_{0}$ for all octet baryons.

The values of $\gamma_0$ are found from the calculation of
$W^{(1)}(\omega)$ via~\cite{TRH98}, \eq \gamma_{0}=\alpha \,
\frac{\partial^{2}}{\partial \omega^{2}}
W^{(1)}(\omega)\bigg|_{\omega=0} \en and below we list the expressions
for $\gamma_{0}$ for all the low-lying octet baryons: \eq
\gamma^p_0&=&\gamma^n_0 \, = \, \frac{\alpha}{\pi^2 F_0^2} \,
\biggl[\frac{(D+F)^{2}}{24}\biggl(\frac{1}{M^2_\pi} +
\frac{1}{M^2_K}\biggr)
+\frac{(D-F)^2}{96 M^2_K}\biggr]\,,\nonumber\\
\gamma^{{\Xi}^{0}}_{0}&=&\frac{\alpha}{\pi^2 F_0^2} \,
\biggl[\frac{(D-F)^2}{48 M^2_\pi}
+\frac{(D+F)^2}{32 M^2_K}\biggr]\,, \nonumber\\
\gamma^{{\Xi}^{-}}_{0}&=&\frac{\alpha}{\pi^2 F_0^2} \,
\biggl[-\frac{5D^2}{288 M^2_K}
+\frac{F^2}{32 M^2_K}
+\frac{(D+F)^2}{96 M^2_K}
+\frac{(D-F)^2}{48 M^2_\pi}\biggr]\,,
\nonumber\\
\gamma^{{\Lambda}}_{0}&=&\frac{\alpha}{\pi^2 F_0^2} \,
[\frac{D^2}{144 M^2_K}
+\frac{F^2}{16  M^2_K}
+\frac{D^2}{72  M^2_\pi}\biggr]\,, \\
\gamma^{{\Sigma}^{0}}_{0}&=&\frac{\alpha}{\pi^2 F_0^2} \,
\biggl[\frac{(D+F)^2}{96 M^2_K}
+\frac{(D-F)^2}{96 M^2_K}
+\frac{F^2}{24 M^2_\pi}\biggr]\,, \nonumber\\
\gamma^{{\Sigma}^{-}}_{0}&=&\frac{\alpha}{\pi^2 F_0^2} \,
\biggl[\frac{(D-F)^2}{48 M^2_K}
+\frac{D^2}{72 M^2_\pi}
+\frac{F^2}{24 M^2_\pi}\biggr]\,,\nonumber\\
\gamma^{{\Sigma}^{+}}_{0}&=&\frac{\alpha}{\pi^2 F_0^2} \,
\biggl[\frac{(D+F)^2}{48 M^2_K} +\frac{D^2}{72 M^2_\pi}
+\frac{F^2}{24 M^2_\pi}\biggr] \,. \nonumber \en We note that in the
nucleon case, when we neglect the kaon loops contributions, we
reproduce the well known result of SU(2) HBChPT~\cite{BER92}.
The other results for spin polarizabilities are new predictions.  In
Table I, the second and third columns give the contribution to
$\gamma_{0}$ from $\pi$ and $\pi+K$ loops, respectively.  In Table I
we also present the results for the nucleon $\gamma_{0}$ obtained in
HBChPT at ${\cal O}(p^3)$~\cite{BER92}, in HBChPT and BChPT at order
${\cal O}(p^4)$~\cite{KBV2000,XJ2000,GCG2000,BER03} and from the
analysis of electroproduction data~\cite{AMS94,DD98one}. For
computation of the polarizabilities, we use $F_0=92 $ MeV, $D=0.8$,
$F=0.46$, $M_{\pi}=139.57$ MeV and $M_{K}=493.65$ MeV.

\section{Conclusions}

We have presented the LO contribution to spin-dependent Compton
scattering in the framework of HBChPT.  In LO HBChPT, these
contributions are all meson loop effects, with no counterterm or
resonance exchange contribution and hence are a test for the chiral
sector of three-flavor QCD.  There exists a small but finite
contribution from kaon loops to $\gamma_{0}$ for the low-lying
octet baryons except the $\Xi^-$ and $\Xi^0$ states. 
Our result for $\gamma_{0}$ in the case of the proton 
and neutron reproduces the results of the LO
calculation of SU(2) HBChPT when kaon loops are not considered and it 
remains to be seen how the
predictions for the other baryons will compare with future
experiments. On the theoretical side, one needs to perform $O(p^4)$
calculations to improve the predictions of the polarizabilities and
to test the convergence of the chiral expansion.  Additional
calculations are also needed to compute $\gamma_{0}$ in the framework of
BChPT with the IDR prescription in order to test the LO and NLO
HBChPT results.  Work in this direction is in progress.

\begin{acknowledgments}

\noindent
One of the authors (KBV) is grateful to BRNS, DAE, India for funding
the project (San. N 2010/37P/18/BRNS/1031 dated 16/08/2010).
He is also thankful to DAAD foundation for awarding the fellowship
(San No A/10/06672 dated 23/04/2010) and to Institut f\"ur Theoretische
Physik, Universit\"at T\"ubingen for warm hospitality.
This research is also part of the Federal Targeted Program
"Scientific and scientific-pedagogical personnel of innovative Russia"
Contract No. 02.740.11.0238.  The work of BRH is supported in part by the 
National Science Foundation under award 
NSF/PHY/0855119.

\end{acknowledgments}

\newpage

\appendix
\section{Feynman rules}
Vertices from
${\cal L}_{\phi \phi}^{(2)}$
\begin{enumerate}
\item Photon-meson coupling: $k_1$ (in-momentum) and $k_2$
(out-momentum) stand either for $\pi$ or for $K$ mesons

\begin{figure}[H]
\begin{center}
\includegraphics[scale=.8]{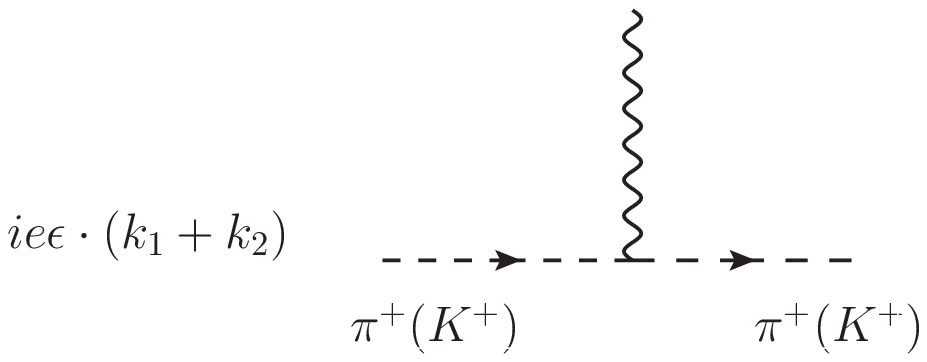}
\end{center}
\end{figure}

\end{enumerate}
\vskip.3in

Vertcies from ${\cal L}^{(1) \, {\rm HBChPT}}_{{\phi}\,B}$
\begin{enumerate}
\setcounter{enumi}{1}

\item Photon-baryon coupling
\begin{figure}[H]
\begin{center}
\includegraphics[scale=.8]{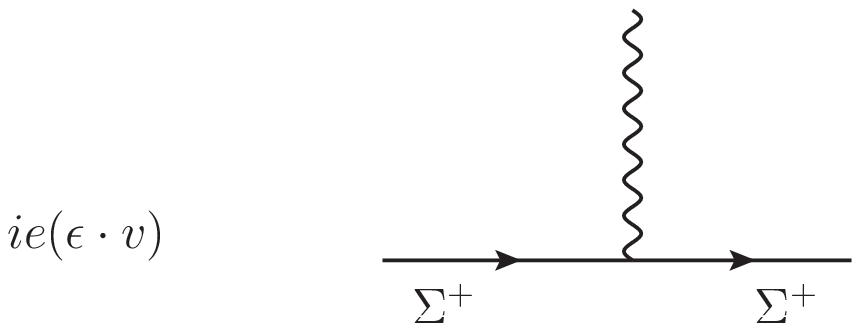}
\end{center}
\end{figure}
\end{enumerate}

Meson-baryon couplings
\begin{enumerate}
\setcounter{enumi}{2}
\item $\pi \Sigma \Sigma$ coupling
\begin{figure}[H]
\begin{center}
\includegraphics[scale=.8]{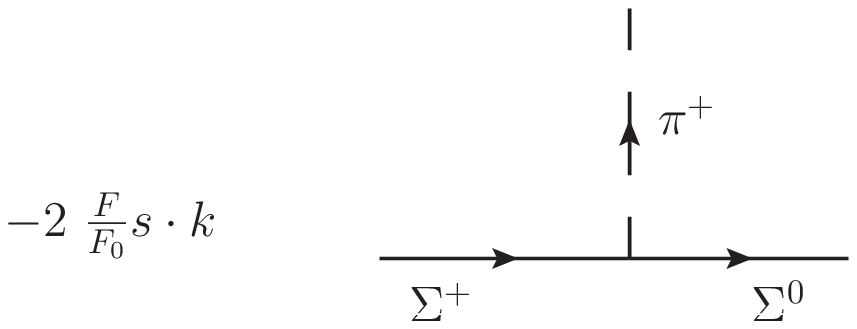}
\end{center}
\end{figure}

\item $K \Sigma \Xi$ coupling
\begin{figure}[H]
\begin{center}
\includegraphics[scale=.8]{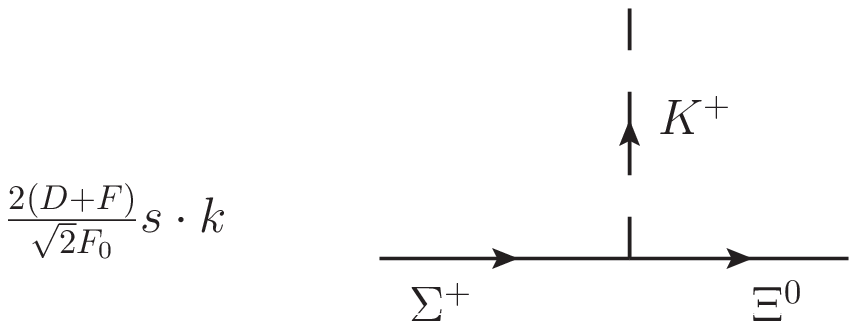}
\end{center}
\end{figure}

\newpage
\item $\pi \Sigma \Lambda$ coupling
\begin{figure}[H]
\begin{center}
\includegraphics[scale=.8]{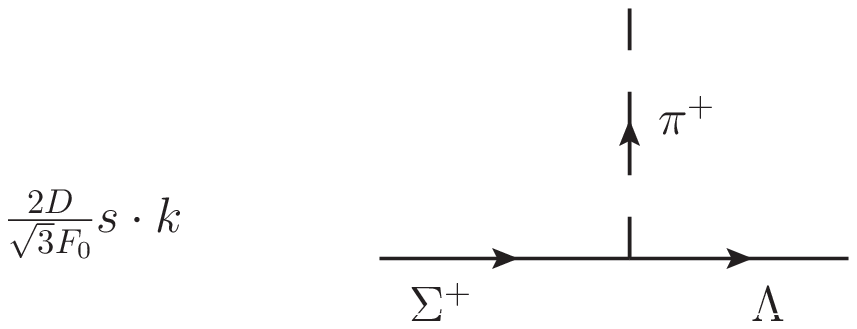}
\end{center}
\end{figure}

\end{enumerate}
Photon-Meson-Baryon couplings
\begin{enumerate}
\setcounter{enumi}{5}
\item $\gamma \pi \Sigma \Sigma$ coupling
\begin{figure}[H]
\begin{center}
\includegraphics[scale=.8]{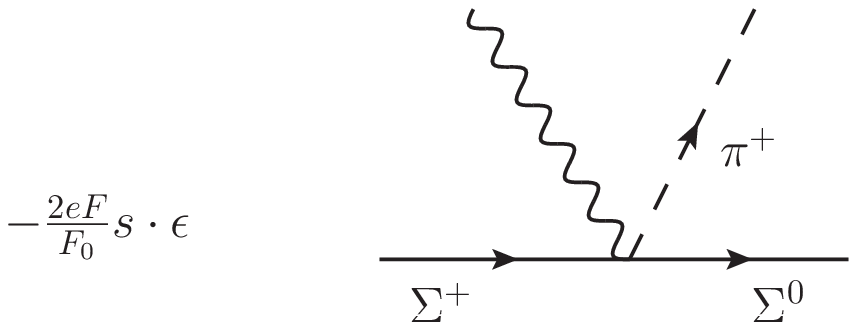}
\end{center}
\end{figure}

\item $\gamma \pi \Sigma \Lambda$ coupling
\begin{figure}[H]
\begin{center}
\includegraphics[scale=.8]{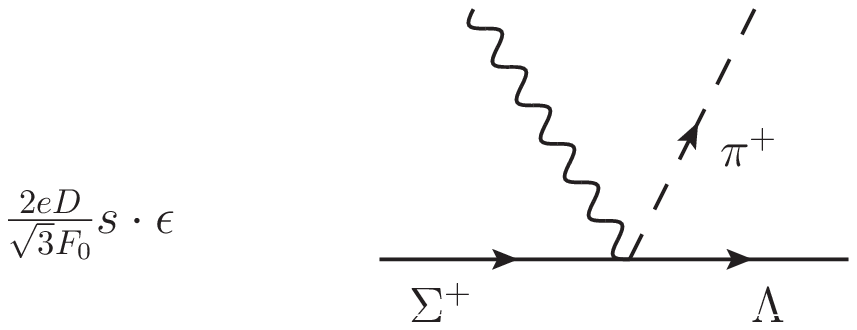}
\end{center}
\end{figure}

\item $\gamma K \Sigma \Xi$ coupling
incoming photon
\begin{figure}[H]
\begin{center}
\includegraphics[scale=.8]{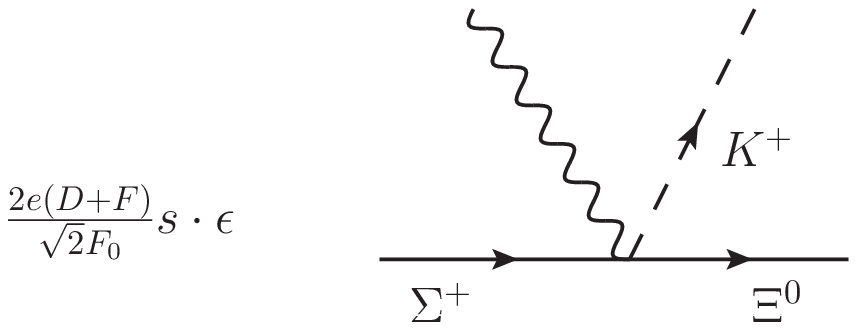}
\end{center}
\caption{\label{feynman_rules}
Feynman rules for evaluating  the $\Sigma^+$
electromagnetic polarizaibilities.}
\end{figure}

\end{enumerate}

\newpage

\section{Loop Integrals}

Here, we have defined all the loop functions which occur in our
calculation and we have given these functions in closed analytical
form as far as possible. In the following all propagators are
understood to have an infinitesimal imaginary part.
The results of the integral are for real photons.
The complete list of integrals can be found in~\cite{BER95}:
\eq
\int \frac{d^dk}{(2\pi)^d i}
\frac{1}{M^2_{P}-k^2} \ = \ \Delta_{P}
\en
where
\eq
\Delta_{P}&=& 2 M^{2}_{P}[L+\frac{1}{16 \pi^{2}}
\log\Big(\frac{4\pi}{\lambda}\Big)+{\cal O}(d-4)]\,, \nonumber\\
L&=&\frac{\lambda^{d-4}}{16 {\pi^{2}}}
[\frac{1}{d-4}+\frac{1}{2}(\gamma_{E}-1-\log(4\pi)]
\en
has a pole at $d=4$. Here P=$ \pi $  or $K$, $\gamma_E=0.557215$ and
$\lambda$ is the scale in dimensional regularization scheme used
in the evaluation of integrals.

The relevant integrals are
\eq
\int \frac{d^dk}{(2\pi)^d i}
\frac{(1, k^\mu, k^\mu k^\nu)}{(v\cdot k-\omega)
[M_P^2-k^2]}&= &
( J^{P}_{0}(\omega), v_{\mu} J^{P}_{1}(\omega),
g^{\mu\nu}J^{P}_{2}(\omega)+v^{\mu} v^{\nu}J^{P}_{3}(\omega) ) \,,
\en
where
\eq
J^{P}_{0}(\omega)&=& -4 L{\omega}+\frac{\omega}{8 {\pi}^2}
(1-2 \log\frac{M_{P}}{\lambda})-
\frac{1}{4{\pi}^2}{\sqrt{M^{2}_{P}-{\omega}^2}}
\arccos\Big(-\frac{\omega}{M_{P}}\Big)+{\cal O}(d-4) \,,\\
J^{P}_{1}(\omega)&=& {\omega} J^{P}_{0}(\omega)+\Delta_{P} \,,\\
J^{P}_{2}(\omega)&=&\frac{1}{d-1}[(M^{2}_{P}-{\omega}^2)
J^{P}_{0}(\omega)- {\omega}{\Delta_{P}} ] \,,\\
J^{P}_{3}(\omega)&=&{\omega} J^{P}_{1 }(\omega)-J^{P}_{2}(\omega) \,.
\en

\section{${\Sigma}^+{\pi}^+(K^{+})$ loops in forward Compton scattering}
\label{sec:TB}

Using the loop integrals  defined in Appendix B, the
${\Sigma^{+}}+ {\pi}^+(K^{+})$ loop diagrams of Fig.~2 can be
written as:
\eq
{\rm Amp}^{{\Sigma}^+{\pi}^+}_{a+a^{\prime}}&=&
C_{1}[S\cdot {\epsilon}^{*},S \cdot{\epsilon}][J^{\pi}_{0}(\omega)
-J^{\pi}_{0}(-\omega)]\\
{\rm Amp}^{{\Sigma}^+{\pi}^+}_{b+c+b^{\prime}+c^{\prime}}&=&
C_{2}[S\cdot {\epsilon}^{*},S \cdot{\epsilon}]
\frac{\partial}{{\partial}M^{2}_{\pi}}\int^{1}_{0}
[J^{\pi}_{2}({\omega}z)-J^{\pi}_{2}(-{\omega}z)] dz\\
{\rm Amp}^{{\Sigma}^+{\pi}^+}_{d+d^{\prime}}&=&
D_{1}[S\cdot {\epsilon}^{*},S \cdot{\epsilon}]
[J^{\pi}_{0}(\omega)-J^{\pi}_{0}(-\omega)] \,, \\
{\rm Amp}^{{\Sigma}^+{\pi}^+}_{e+f+e^{\prime}+f^{\prime}}&=&
D_{2}[S\cdot {\epsilon}^{*},S \cdot{\epsilon}]
\frac{\partial}{{\partial}M^{2}_{\pi}}
\int^{1}_{0}[J^{\pi}_{2}({\omega}z)-J^{\pi}_{2}(-{\omega}z)] dz \,, \\
{\rm Amp}^{{\Sigma}^+{K}^+}_{g+g^{\prime}}&=& E_{1}[S\cdot {\epsilon}^{*},S
\cdot{\epsilon}][J^{K}_{0}(\omega)-J^{K}_{0}(-\omega)]\, \\
{\rm Amp}^{{\Sigma}^+{K}^+}_{h+i+h^{\prime}+i^{\prime}}&=&
E_{2}[S\cdot {\epsilon}^{*},S \cdot{\epsilon}]
\frac{\partial}{{\partial}M^{2}_{K}}\int^{1}_{0}
[J^{K}_{2}({\omega}z)-J^{K}_{2}(-{\omega}z)] dz \,,
\en
where
\eq
C_{1}&=& 2i \, \Big(\frac{e \, F}{F_0}\Big)^2\,, \quad
C_{2} \ = \ -8i \, \Big(\frac{e \, F}{F_0}\Big)^2\,, \\
D_{1}&=& \frac{2i}{3} \, \Big(\frac{e \, D}{F_0}\Big)^2\,, \quad
D_{2} \ = \ - \frac{8i}{3} \, \Big(\frac{e \, D}{F_0}\Big)^2\,, \\
E_{1}&=& i \, \Big(\frac{e \, (D+F}{F_0}\Big)^2\,, \quad
E_{2} \ = \  - 4i \, \Big(\frac{e \, (D+F}{F_0}\Big)^2\,.
\en

\begin{figure}
\begin{center}
\includegraphics[scale=.8]{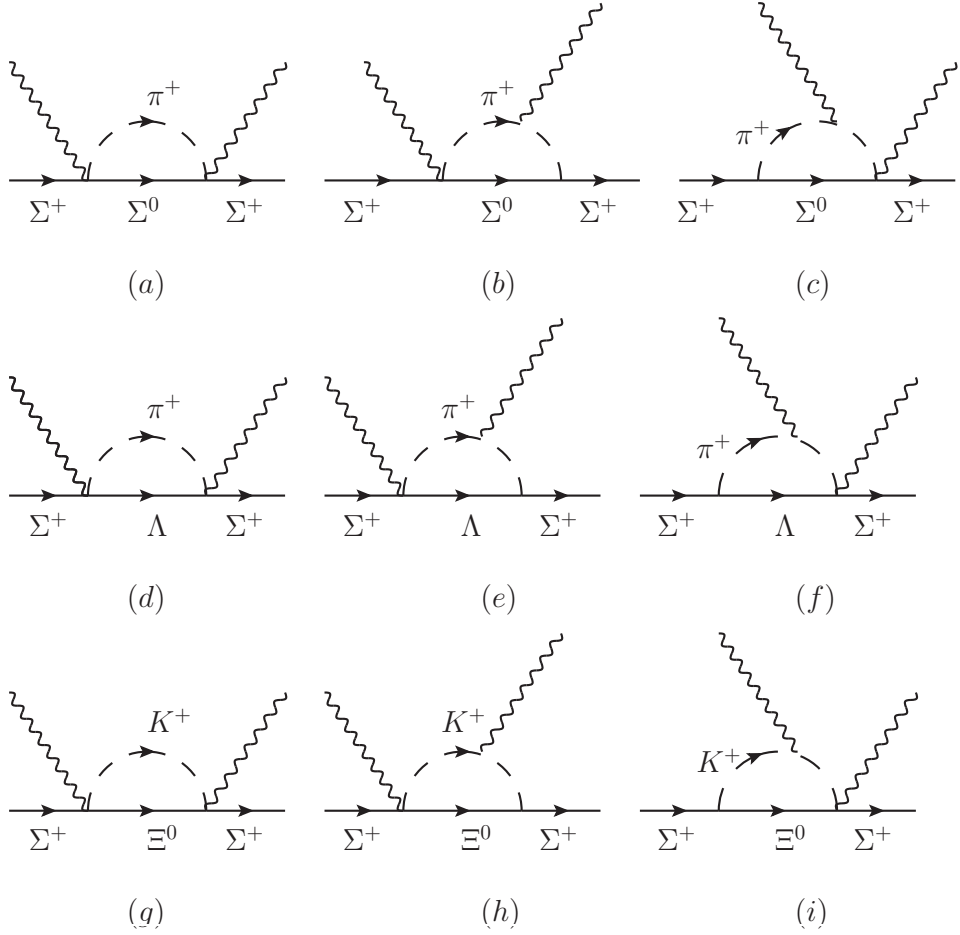}
\end{center}
\caption{\label{fig1} The one loop diagrams contributing
to forward Compton scattering of $\Sigma^+$ $\pi^+$ $(K^+)$ at $O(p^3)$.
Crossed diagrams are not shown.}
\end{figure}

\newpage

\begin{table}
\caption{\label{Tab:sing-P} The forward spin polarizablity $\gamma_{0}$
of octet baryons (in units of $10^{-4}$ fm$^{4}$) }

\hspace*{-1cm}
\begin{tabular}{|c|c|c|c|c|c|}
\hline
Baryon& Our results at $O(p^{3})$
& Our results at $O(p^{3})$&$ O(p^{3})\,$
&$ O(p^{4})\,$& \\

&with $\pi$ loops&with  $\pi$ and $K$ loops
& HBChPT~\cite{BER92}& HBChPT and BChPT
& Electroproduction data\\

\hline
$p$& 4.50 & 4.86 & 4.5 & $4.5-(6.9+1.5)$~\cite{KBV2000,XJ2000,GCG2000},\
     4.64~\cite{BER03} & $-$1.3~\cite{AMS94}, \
$-$0.6~\cite{DD98one},\ \\
& & & & & $-0.58\pm 0.2$~\cite{SCH2011}\\
\hline
$n$& 4.50 & 4.86 & 4.5 & $4.5-(6.9-1.5)$~\cite{KBV2000,XJ2000,GCG2000},\
   1.82~\cite{BER03}& $-$0.4~\cite{AMS94},\ \\
& & & & & $- 0.38\pm 0.22$~\cite{SCH2011}\\
\hline
$\Sigma^{+}$& 1.20 & 1.38 &&&\\
\hline
$\Sigma^{0}$& 0.60 & 0.70 &&&\\
\hline
$\Sigma^{-}$& 1.20 & 1.22 &&&\\
\hline
$\Lambda$& 0.60 & 0.70 &&&\\
\hline
$\Xi^{-}$& 0.16 & 0.26 &&&\\
\hline
$\Xi^{0}$& 0.16 & 0.43 &&&\\
\hline

\end{tabular}
\end{table}

\end{document}